\documentclass[sn-mathphys-ay]{sn-jnl}

\usepackage{geometry}
\usepackage{graphicx}%
\usepackage{multirow}%
\usepackage{amsmath,amssymb,amsfonts}%
\usepackage{amsthm}%
\usepackage{mathrsfs}%
\usepackage[title]{appendix}%
\usepackage{xcolor}%
\usepackage{textcomp}%
\usepackage{manyfoot}%
\usepackage{booktabs}%
\usepackage{algorithm}%
\usepackage{algorithmicx}%
\usepackage{algpseudocode}%
\usepackage{listings}%
\graphicspath{{figures/}}
\usepackage{upgreek}
\usepackage[size=small]{subcaption}
\usepackage[separate-uncertainty=true,multi-part-units=single]{siunitx}
\usepackage{gensymb}
\usepackage{cleveref}
\usepackage{blindtext}
\usepackage{verbatim}
\usepackage[colorinlistoftodos,prependcaption,textsize=tiny]{todonotes}

\newcommand\itblue[1]{\textcolor{blue}{\textit{#1}}}

\colorlet{Reviewer1}{orange}
\colorlet{Reviewer2}{cyan}

\theoremstyle{thmstyleone}%
\theoremstyle{thmstyletwo}%

\theoremstyle{thmstylethree}%

\begin{document}

\title[Article Title]{Fluid wetting and penetration characteristics in T-shaped microchannels}


\author[1,3]{\fnm{Huijie} \sur{Zhang}}

\author*[2]{\fnm{Anja} \sur{Lippert}}\email{Anja.Lippert@bosch.com}

\author[2]{\fnm{Ronny} \sur{Leonhardt}}

\author[2]{\fnm{Tobias} \sur{Tolle}}

\author[2,3]{\fnm{Luise} \sur{Nagel}}


\author*[3]{\fnm{Tomislav} \sur{Mari\'c}}\email{maric@mma.tu-darmstadt.de}

\affil[1]{\orgdiv{Mobility Electronics}, \orgname{Robert Bosch GmbH}, \orgaddress{\street{Markwiesenstrasse 46}, \city{Reutlingen}, \postcode{72770}, \country{Germany}}}

\affil[2]{\orgdiv{Corporate Research}, \orgname{Robert Bosch GmbH}, \orgaddress{\street{Robert-Bosch-Campus 1}, \city{Renningen}, \postcode{71272}, \country{Germany}}}

\affil[3]{\orgdiv{Mathematical Modeling and Analysis}, \orgname{Technical University of Darmstadt}, \orgaddress{\street{Peter-Gruenberg-Strasse 10}, \city{Darmstadt}, \postcode{64287}, \country{Germany}}}


\abstract{\itblue{This is the author manuscript of an article \cite{zhang2024_published} published in Experiments in Fluids, available at \url{https://rdcu.be/dZM07}. Please cite the published journal article \cite{zhang2024_published} when refering to this manuscript.} A thorough understanding of media tightness in automotive electronics is crucial for ensuring more reliable and compact product designs, ultimately improving product quality. Concerning the fundamental characteristics of fluid leakage issues, the dynamic wetting and penetration behavior on small scales is of special interest and importance. In this work, four T-shaped microchannels with one inlet and two outlets are experimentally investigated in terms of contact angle dynamics and interface movement over time, generating novel insight into the wetting mechanisms and fluid distribution. With a main channel width of $\SI{1}{mm}$, a crevice width of $w = \SI{0.3}{mm}, \SI{0.4}{mm}$ and a rounding edge radius of $r = \SI{0.1}{mm}, \SI{0.2}{mm}$, the geometrical effects on the fluid penetration depth in the crevice and the interface edge pinning effect are analyzed quantitatively using an automated image processing procedure. It is found that the measured dynamic contact angles in all parts can be well described by molecular kinetic theory using local contact line velocities, even with local surface effects and abrupt geometry changes. Moreover, a smaller crevice width, a sharper edge and a larger flow velocity tend to enhance the interface pinning effect and prevent fluid penetration into the crevice. The rounding radius has a more significant effect on the interface pinning compared with crevice width. The experimental data and image processing algorithm are made publicly available.}




\maketitle

\section{Introduction}\label{sec:01_introduction}

Power electronics play an increasingly vital role, not only in the automotive industry with electrified vehicles, but also in modern energy systems, providing, for example, advantages with respect to power density and efficiency. With rapid development and growing demand of power electronics in industrial applications, yielding a better and more reliable product design has become essential (\cite{Wang2021}, \cite{YANG20241387}). Among all the considered failure models, the fluid penetration and accumulation through unforeseen microfractures along sealing joints or other connection positions have attracted more attention due to its complexity and unpredictability, as reported by \cite{Ciprian2009}, \cite{Hygum2015}. In particular, while significant progress has been made in the wetting characteristics on small scales, there are still challenges in achieving a comprehensive understanding and predictive modeling concerning geometrically complex leakage scenarios.

Wetting phenomena concern the moving interface between liquids and solids (\cite{deGennes1985}), which are present in numerous applications and is under active research analytically, experimentally, and numerically, as recently reviewed by \cite{ZHANG2023102861}. Contrary to well-studied static wetting, dynamic wetting with contact line (CL) motion is still an active research field, especially in the context of penetration behavior. With extensive research being undertaken in the past decades, several analytical models have been developed to describe CL dynamics. The hydrodynamic theory (HDT) (\cite{Voinov1976}, \cite{Cox1986}) concentrates on the macroscopic motion of the fluid interface on the solid substrate, while the molecular kinetic theory (MKT) focuses on the CL friction dissipation in the vicinity of the contact zone on the molecular scale. A combined molecular-hydrodynamic model is proposed by \cite{Petro1992}, \cite{BROCHARDWYART19921} and \cite{YANG202021} as a result of the limitations of both models. Furthermore, the interface formation model (IFM) (\cite{SHIKHMURZAEV1993589}) is added to the analytical solution. Widely used empirical models based on experimental findings are given by \cite{JIANG197974} and \cite{Bracke1989}.

The T-shaped channel is essential in a wide range of scientific and industrial applications, especially for the phase separation and mixing processes, as reviewed by \cite{YANG2019109895} and \cite{LU2022112742}. The gas-liquid two-phase separation mechanism in T-channel is widely utilized due to its simplicity in design and manufacturing, as well as its effective separation capability (\cite{YANG2024257}). Speaking of separated two-phase flow in a T-shaped microchannel, the focus of previously published works has been mainly limited to droplet formation (\cite{Sivasamy2011}, \cite{Wehking2014}) and breakup mechanisms (\cite{Leshansky2009}). However, only few studies have been carried out on addressing the fluid penetration characteristics in a branching-off channel, which has a much smaller width than the inlet channel. The interface pinning effect at the dividing position remains still largely unexplored. An in-depth investigation of this effect expands our understanding of two-phase flow processes in various applications.

While extensive experimental and numerical studies are reported for more controlled wetting dynamics, such as capillary rise (\cite{Qur1997InertialC}, \cite{Vega2005}), spreading (\cite{ArjmandiTash2017KineticsOW}, \cite{HAN2021110190}) and impinging droplet (\cite{Sikalo2005}, \cite{YANG2021127634}), less attention has been paid to the forced wetting through microchannels, which is of considerable and practical importance in the context of industrial sealing circumstances. In a recent work, \cite{Zhang2024} experimentally investigated the wetting dynamics of two working fluids through four distinct curved microchannels. The MKT model is reported to be superior to other theoretical approaches in describing the CL movement with local surface effects.

This work reports a comprehensive experimental study on the CL dynamics and penetration characteristics at small capillary numbers $Ca$ $(10^{-6} - 10^{-4})$ through four T-shaped microchannels, distinguished by the crevice width $w = \SI{0.3}{mm}, \SI{0.4}{mm}$ and rounding radius $r = \SI{0.1}{mm}, \SI{0.2}{mm}$. A robust automated image processing method is proposed and utilized to evaluate the dynamic contact angle and interface displacement evolution. The comparison between the measured dynamic contact angles and MKT shows great agreement with physically reasonable fitting parameters, confirming the reliability and reproducibility of the experimental results and image processing method. In addition, the geometrical variations of the channel are found to have an intuitive and significant influence on the penetration depth and edge pinning effect, providing new insights into reliable product design concerning sealing performance.

\section{Methodology}\label{sec:02_methodology}

\subsection{Materials and samples}\label{subsec:2.1_samples}

\begin{figure}[!htb]
	\centering
	\includegraphics[width=\linewidth]{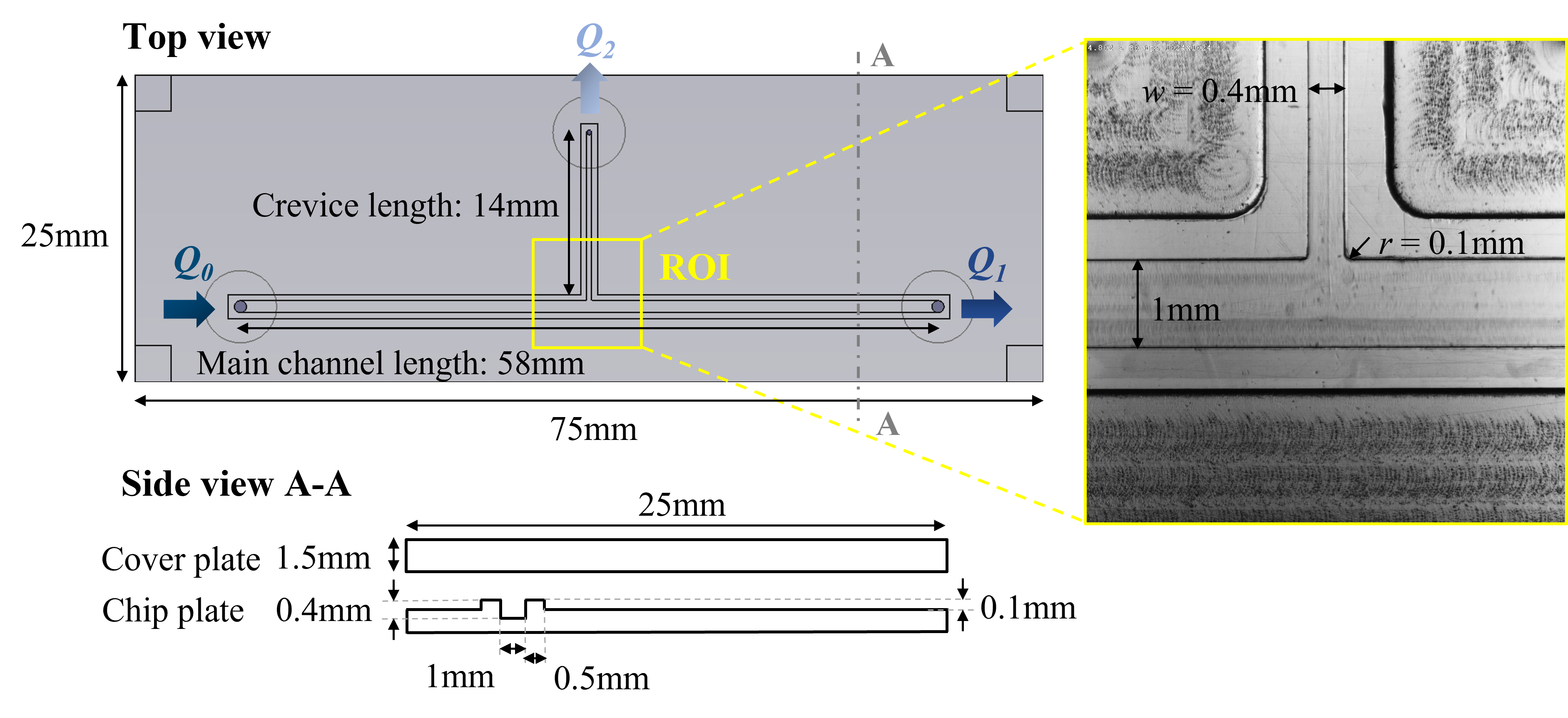}
	\caption{Chip plate with T-shaped microchannel (crevice width $w=\SI{0.4}{mm}$, rounding radius $r=\SI{0.1}{mm}$). Top: top view with region of interest (ROI). Bottom: side view A-A with cover plate.}
	\label{fig:2.1_sample}
\end{figure}

For the experimental study, chip plates as Topas (a cyclo-olefin copolymer) microscopy slides (Microfluidic Chipshop) are customized with T-shaped open microchannels by micromilling and employed as test samples. One more smooth Topas cover plate is used to enclose the channel by a preload force from the clamping assembly. The geometry and dimensions of the test sample are presented in \cref{fig:2.1_sample}. The manufactured channels have rectangular cross-sections, which are enclosed by side walls (width \SI{0.5}{mm}, height \SI{0.1}{mm}). No additional glue or sealing materials are used to avoid contamination and ensure the same wetting properties on all walls. Although the fluid infiltration into the joint between the top of the side wall and the cover plate cannot be entirely prevented, as a result of the surface roughness of $Rz=0.08$ of the original material, the theoretical maximum of the wall-sealing caused leakage is approximately 0.05\%. The theoretical maximum fluid leakage is calculate by assuming that the sealing wall is entirely filled with fluid leakage. For the sample with a crevice width $w=\SI{0.4}{mm}$, the theoretical maximum fluid leakage is determined by calculating the ratio between theoretical flow velocity $U_{\text{theo}}$ and the real flow velocity $U_{\text{real}}$, as follows:
\begin{equation}
\frac{U_{\text{real}}}{U_{\text{theo}}} = \frac{A_{\text{theo}}}{A_{\text{real}}} = \frac{0.4 \times 0.4}{0.4 \times 0.4 + 1 \times 0.08 \times 0.001} = 0.0005.
\end{equation}
Hence, the mass loss from leakage is negligible. The impact of the leakage on the CL dynamics is also negligible, given that this small amount of leakage is concentrated around the contact point at the edge between the side wall and the cover plate, while the CL is a curve that spans the entire width of side edge and the cover plate. With a well-defined clamping assembly regarding force and position, fluid infiltration is rarely detected in most cases. However, it can be observed in some cases at low flow rates and is discussed in detail in \cref{subsec:3.2_dynCA}. It is important to note that the channels are not intended to be sealed permanently, as they need to be re-opened for further investigations, such as cleaning procedures or microscopy analysis. Other bonding methods, such as glue, double-sided tape, and solvent bonding, were tested but ultimately excluded since they could introduce unknown variables into the system by altering both the channel geometry and the surface properties of the channel walls. Therefore, closing the channels by incorporating a sealing wall was chosen as a more controlled and reliable approach.


As shown schematically in \cref{fig:2.1_sample}, three connector ports are present on the chip plate: one inlet on the left side of the main channel, two outlets on the top of the crevice, and the right side of the main channel. All samples have a constant channel depth of \SI{0.4}{mm} and a main channel width of \SI{1}{mm}. They differ in crevice width $w = \SI{0.3}{mm}, \SI{0.4}{mm}$ and rounding edge radius $r = \SI{0.1}{mm}, \SI{0.2}{mm}$, resulting in four sample variations listed in \cref{tab:2.1_sample}. The visualization region of interest (ROI) is marked as a yellow square in \cref{fig:2.1_sample}.

As working fluid, de-ionized water is used and the physical properties are internally determined and given in \cref{tab:2.2_fluid}.

\begin{table}[!htb]
	\centering
	\begin{minipage}[b]{0.5\textwidth}
		\centering
		\begin{tabular}{c c c}
			\toprule[1.5pt]
			Sample & $w$ (\SI{}{mm}) & $r$ (\SI{}{mm}) \\
			\midrule[0.5pt]
			w0.4\_r0.1 & 0.4 & 0.1 \\
			w0.4\_r0.2 & 0.4 & 0.2 \\
			w0.3\_r0.1 & 0.3 & 0.1 \\
			w0.3\_r0.2 & 0.3 & 0.2 \\
			\bottomrule[1.5pt]
		\end{tabular}
		\caption{The test sample variations with crevice width $w$, rounding edge radius $r$.}
		\label{tab:2.1_sample}
	\end{minipage}
	\hfill
	\begin{minipage}[b]{0.49\textwidth}
		\centering
		\begin{tabular}{c c c c}
			\toprule[1.5pt]
			Working fluid & density & dyn. viscosity & surface tension  \\
			@ \SI{22}{\celsius} & $\rho\ (\SI{}{kg/m^3})$ & $\mu\ (\SI{}{Pa \cdot s})$ & $\sigma\ (\SI{}{N/m})$ \\
			\midrule[0.5pt]
			Water & 998.03 & 1.000e-3 & 72.74e-3 \\
			\bottomrule[1.5pt]
		\end{tabular}
		\caption{Physical properties of the working fluid and surface tension with air.}
		\label{tab:2.2_fluid}
	\end{minipage}
\end{table}
\vspace{\belowdisplayskip}

\subsection{Experimental setup and procedure}\label{subsec:2.2_setup}

\begin{figure}[!htb]
	\centering
	\includegraphics[width=0.9\linewidth]{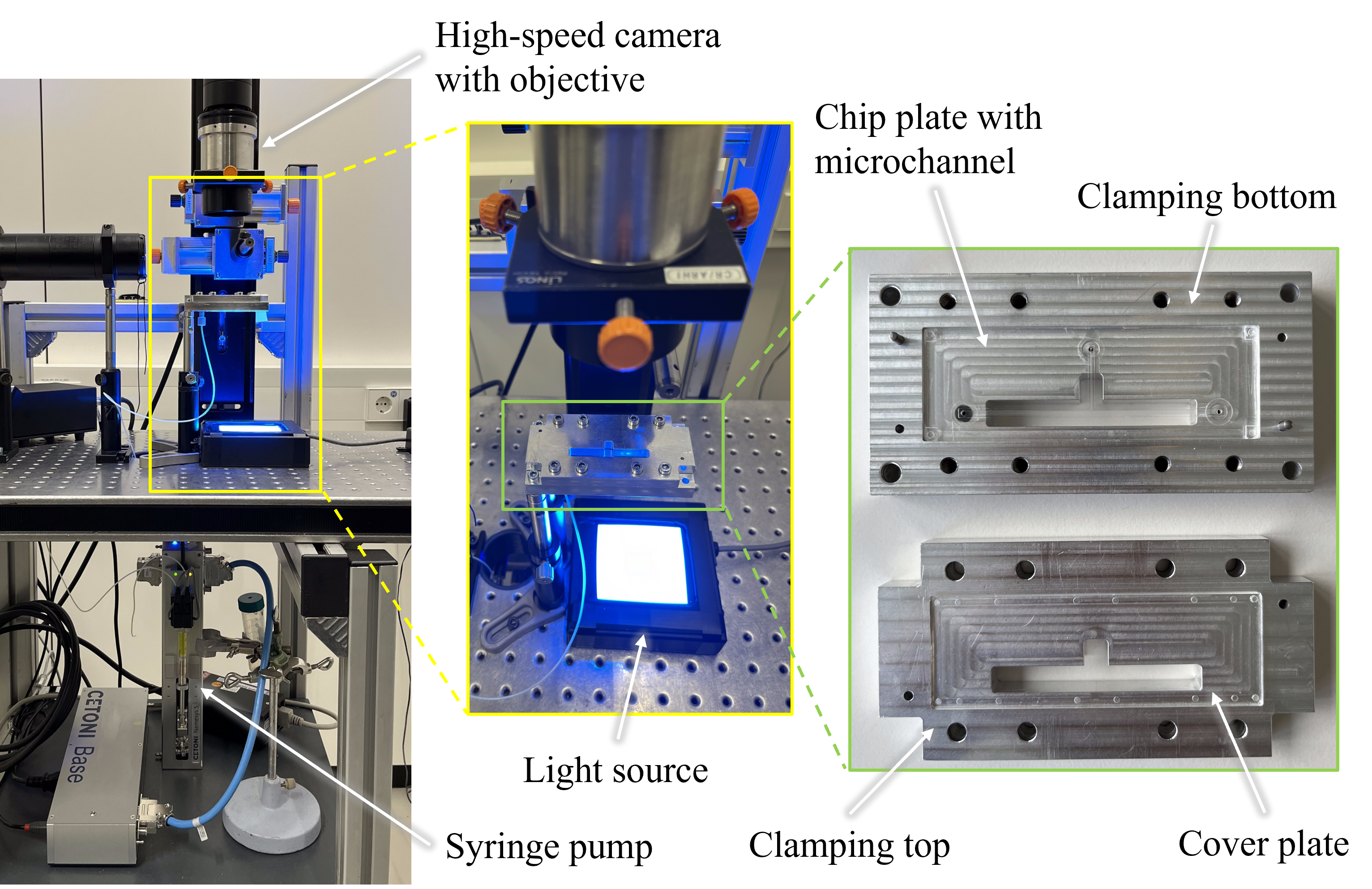}
	\caption{Experimental setup. Left: test bench. Middle: zoom-in view. Right: chip plate and cover plate with clamping assembly.}
	\label{fig:2.2_setup}
\end{figure}

The experimental setup is shown in \cref{fig:2.2_setup} with the test bench on the left side, a zoom-in view in the middle and the chips with clamping assembly on the right side. The clamping assembly consists of two parts to ensure a plane and reproducible force: the top for placing the cover plate and the bottom for the chip plate. Both parts of the clamping assembly are connected with alignment pins and eight equally distributed screws. The clamping force is controlled using a torque wrench. A syringe pump (CETONI Nemesys S) is used to control the inlet fluid flow and the moving interface is captured optically by a high-speed camera (NX4-S3, LDT). The emitted light from a light source is transmitted through a visualization window of the clamping assembly and the transparent chips directly to the camera.

The ROI is visualized by 1024$\times$1024 pixels with an image resolution of $\Delta x=\SI{5.4e-6}{m}$. The investigated volumetric flow rates $Q_0$ and corresponding recorded frame rates are listed in \cref{tab:2.3_opticalParameter}. The data acquisition is realized with MotionStudio.

\begin{table}[!htb]
	\centering
	\begin{tabular}{c c c c}
		\toprule[1.5pt]
		$Q_0 \ (\SI{}{ml/s})$ & $U \ (\SI{}{mm/s})$ & $Ca$ & Frame rate (fps) \\
		\midrule[0.5pt]
			0.0002 & 0.5 & $\SI{6.88e-6}{}$ & 30 \\
			0.0004 & 1 & $\SI{1.37e-5}{}$ & 60 \\
			0.002 & 5 & $\SI{6.88e-5}{}$ & 100 \\
			0.003, 0.004 & 7.5, 10 & $\SI{1.03e-4}{}$, $\SI{1.37e-4}{}$ & 200 \\
			0.006 & 15 & $\SI{2.06e-4}{}$ & 400 \\
		\bottomrule[1.5pt]
	\end{tabular}
	\caption{Investigated volumetric flow rates $Q_0$, flow velocities $U$, capillary numbers $Ca$ in experiments and the corresponding optical frame rates.}
	\label{tab:2.3_opticalParameter}
\end{table}

The static contact angle (CA) of the sample material is firstly determined as $\theta_0=93 \pm 5 \degree$ by placing a water droplet with a volume of $V=\SI{1}{\mu L}$ on the chip surface after micromilling. For all operation points, the working fluid is pumped through the inlet port into the main channel with a constant volumetric flow rate $Q_0$ and leaves the system via the outlet in the crevice with $Q_2$ and the outlet in the main channel with $Q_1$, as denoted in \cref{fig:2.1_sample}. After each test, the clamping assembly is opened and the chips are cleaned with water and dried with compressed air. Each test is repeated three times to ensure the reproducibility and accuracy of the results. The image recording is manually started before the interface enters the ROI and stopped after the interface leaves the crevice in the ROI. After the image acquisition, the image series are processed for the CA measurement and interface movement over time via an automated workflow (\cref{subsec:2.3_imageProcessing}).

\subsection{Automated image processing}\label{subsec:2.3_imageProcessing}

\begin{figure}
	\centering
	\def\svgwidth{\linewidth}
	\input{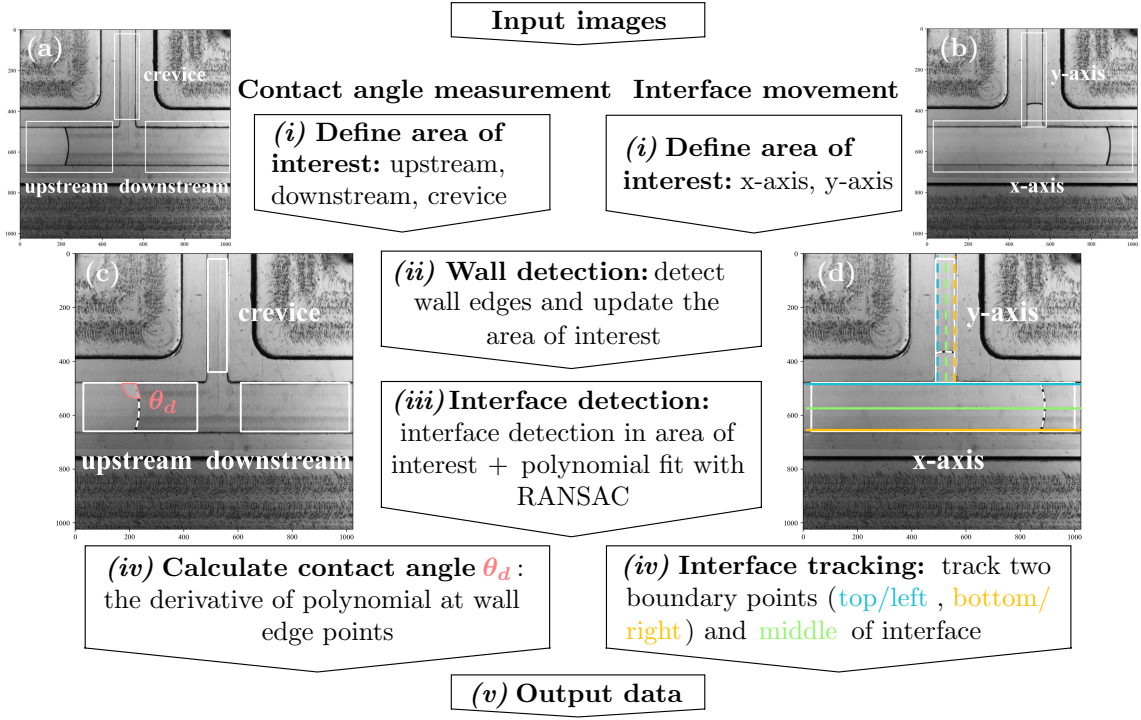}
	\caption{Automated image processing workflow.}
	\label{fig:2.3_imageFlowChart}
\end{figure}

As mentioned in \cref{subsec:2.2_setup}, the dynamic CA and temporal evolution of interface displacement are derived from the recorded images. For accurate and standardized data processing, an automated workflow is proposed as displayed in \cref{fig:2.3_imageFlowChart} and publicly available on GitHub (\cite{github}) and TUDataLib (\cite{TUDatalib}). The procedure is explained hereafter in four steps and validated in \cref{subsec:3.1_flowRate}. 

\paragraph{(i) Define area of interest}
Due to the T-shaped geometry, the area of interest for further processing needs to be determined first. For the CA measurement, three areas named ``upstream", ``downstream" and ``crevice" excluding the T-junction, are chosen for the stable and symmetrical interface shape (\cref{fig:2.3_imageFlowChart}(a)). For the interface tracking, different areas of interest are selected. To enable continuous tracking along the main channel and the crevice, the areas ``x-axis" and ``y-axis" are chosen (\cref{fig:2.3_imageFlowChart}(b)). It is worth mentioning that the exact position of the clamping assembly might vary slightly in each sample change due to the tolerance of holder screws. Therefore, the preliminary selection of all areas of interest is made slightly larger than the expected dimension as shown in \cref{fig:2.3_imageFlowChart}(a) and (b).

\paragraph{(ii) Wall detection}
With the preliminary areas of interest being chosen, the wall edges are detected on the first image of the series as the first two darkest pixels along ten vertical lines in horizontal areas or ten horizontal lines in vertical areas. The coordinates' median value of the sampled darkest pixels are considered as reference wall edges and then updated as final areas of interest (\cref{fig:2.3_imageFlowChart}(c) and (d)) for the rest of the images, where the interface is detected for further analysis.

\paragraph{(iii) Interface detection}
Within the final areas of interest, the interface is considered to contain the darkest pixels and can be approximated with a polynomial fit. The Random Sample Consensus (RANSAC) developed by \cite{FISCHLER1987726} is utilized using \cite{ransac} library for performing the polynomial fit and filtering outliers, for example, scratches or imprints on the sample might be detected as outliers. This approach follows the methodology described by \cite{Nagel2024}, where RANSAC was successfully applied to achieve robust polynomial fits in the presence of outliers. The fitted interface is marked as a white dashed curve in \cref{fig:2.3_imageFlowChart}(c) and (d). A polynomial degree of three is shown to be very robust and accurate for the interface approximation, as confirmed by the low deviation of the statistical results and also by \cite{Nagel2024}.

\paragraph{(iv) Calculate contact angle}
After the interface is acquired, the CA is calculated in a global approach with the derivative of the polynomial at intersection points between walls and polynomial, as displayed in \cref{fig:2.3_imageFlowChart}(c).

\paragraph{(iv) Interface tracking}
Two boundary points and the middle point of the interface, as marked in \cref{fig:2.3_imageFlowChart}(d), are tracked to quantify the interface movement over time. The two boundary points are chosen to be three pixels away from the wall edges to exclude the local wall effects. Once the coordinates of the three points are determined, the temporal evolution of the interface displacement can be computed using the image resolution and optical frame rate information.

\paragraph{(v) Output data}
As output, the CA measurements in ``upstream", ``downstream", ``crevice", and the interface displacement in ``x-axis", ``y-axis" areas are saved as CSV files and can be used for further data visualization.

\vspace{\belowdisplayskip}
The proposed image processing method provides an automated, robust and transferable workflow for deriving the CA and tracking the interface motion. There are four adjustable parameters in the algorithm: the threshold value for classifying the pixel values \texttt{thres}, the number of lines for the channel wall detection \texttt{n\_lines}, the polynomial degree for the interface fitting \texttt{poly\_degree} and the number of offset pixels to the wall edges \texttt{n\_pixels} for the interface tracking. All four parameters are confirmed to be non-sensitive and kept the same for all data sets as \texttt{thres} = 50, \texttt{n\_lines} = 10, \texttt{poly\_degree} = 3 and \texttt{n\_pixels} = 3.

\section{Results and discussion}\label{sec:03_results}

\subsection{Flow rate distribution}\label{subsec:3.1_flowRate}

\begin{figure}[!htb]
	\centering
	\begin{subfigure}[b]{.66\linewidth}
		\centering
		\includegraphics[width=\linewidth]{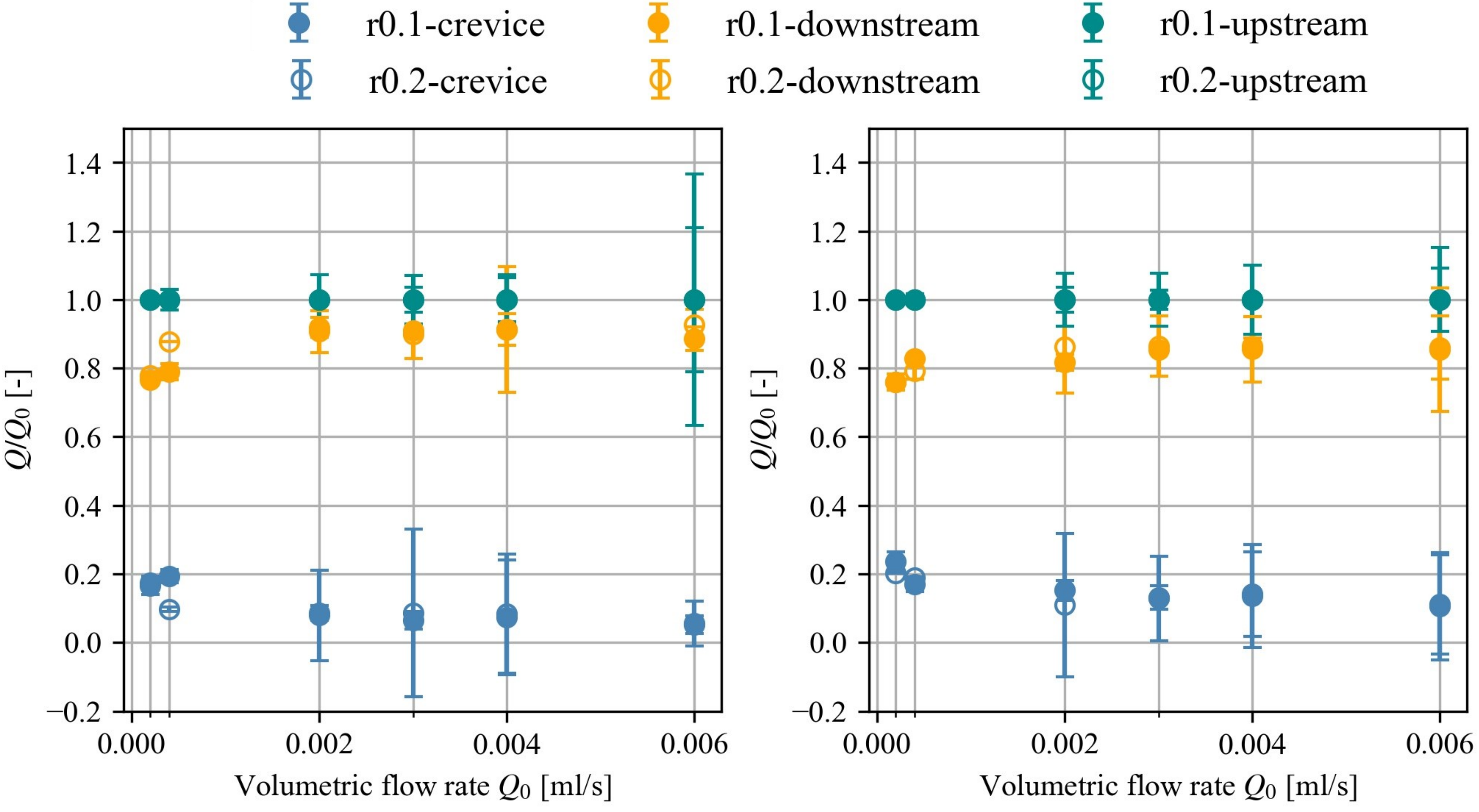}
		\caption{Left: $w=\SI{0.4}{mm}$. Right: $w=\SI{0.3}{mm}$}
		\label{fig:3.1.1_flowRateDistribution}
	\end{subfigure}
	\hfill
	\begin{subfigure}[b]{.33\linewidth}
		\centering
		\includegraphics[width=\linewidth]{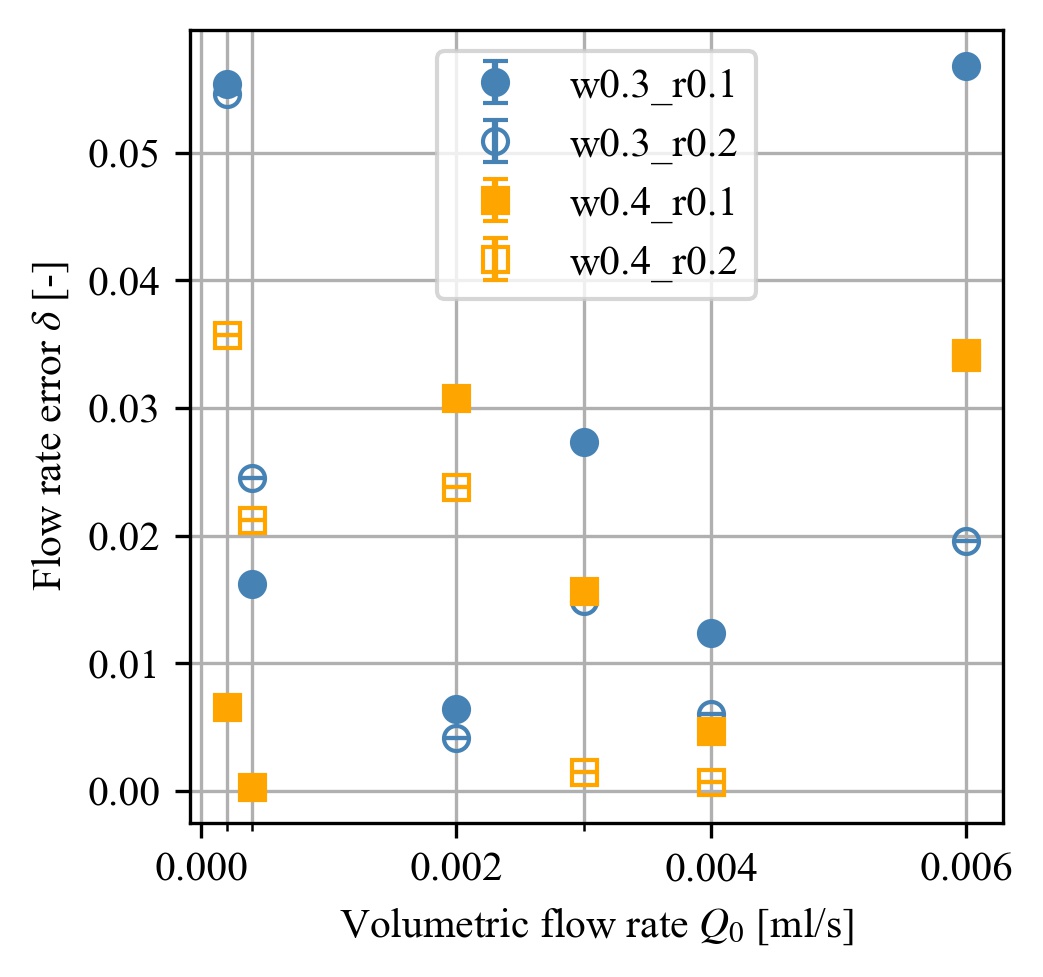}
		\caption{Flow rate error $\delta$}
		\label{fig:3.1.2_flowRateError}
	\end{subfigure}
	\caption{Flow rate distribution $Q/Q_0$ with $Q$ as the volumetric flow rate in the ``upstream", ``downstream" and ``crevice" region and the flow rate error $\delta$ (\cref{eq:flowRateError}).}
	\label{fig:3.1_flowRate}
\end{figure}

For the validation of the image processing method, the volumetric flow rate distribution in the T-shaped channel is evaluated, and the results are presented in \cref{fig:3.1.1_flowRateDistribution}. $Q$ denotes the volumetric flow rate in the ``upstream", ``downstream" and ``crevice" region and is calculated with the area of the channel cross-section $A$ and the interface velocity $U$ as $Q=A \times U$. It can be seen in \cref{fig:3.1.1_flowRateDistribution} that with smaller crevice width $w$, the amount of the delivered fluid into the crevice is decreasing, while it is increasing along the main channel. To quantify the validity and accuracy of the image processing workflow, an error indicator $\delta$ based on the mass conservation is introduced as
\begin{equation}
\delta=\frac{Q_0-Q_1-Q_2}{Q_0},
\label{eq:flowRateError}
\end{equation}
where $Q_1$ and $Q_2$ are the flow rates as denoted in \cref{fig:2.1_sample}. The low error values in \cref{fig:3.1.2_flowRateError} confirm that the experimental results are highly reliable and reproducible. The automated image processing method accurately captures flow velocities so that also the calculated volumetric flow rates, with minimal deviation from theoretical expectations.

\subsection{Dynamic contact angle}\label{subsec:3.2_dynCA}

\begin{figure}[!htb]
	\centering
	\begin{subfigure}[t]{.5\linewidth}
		\centering
		\includegraphics[width=\linewidth]{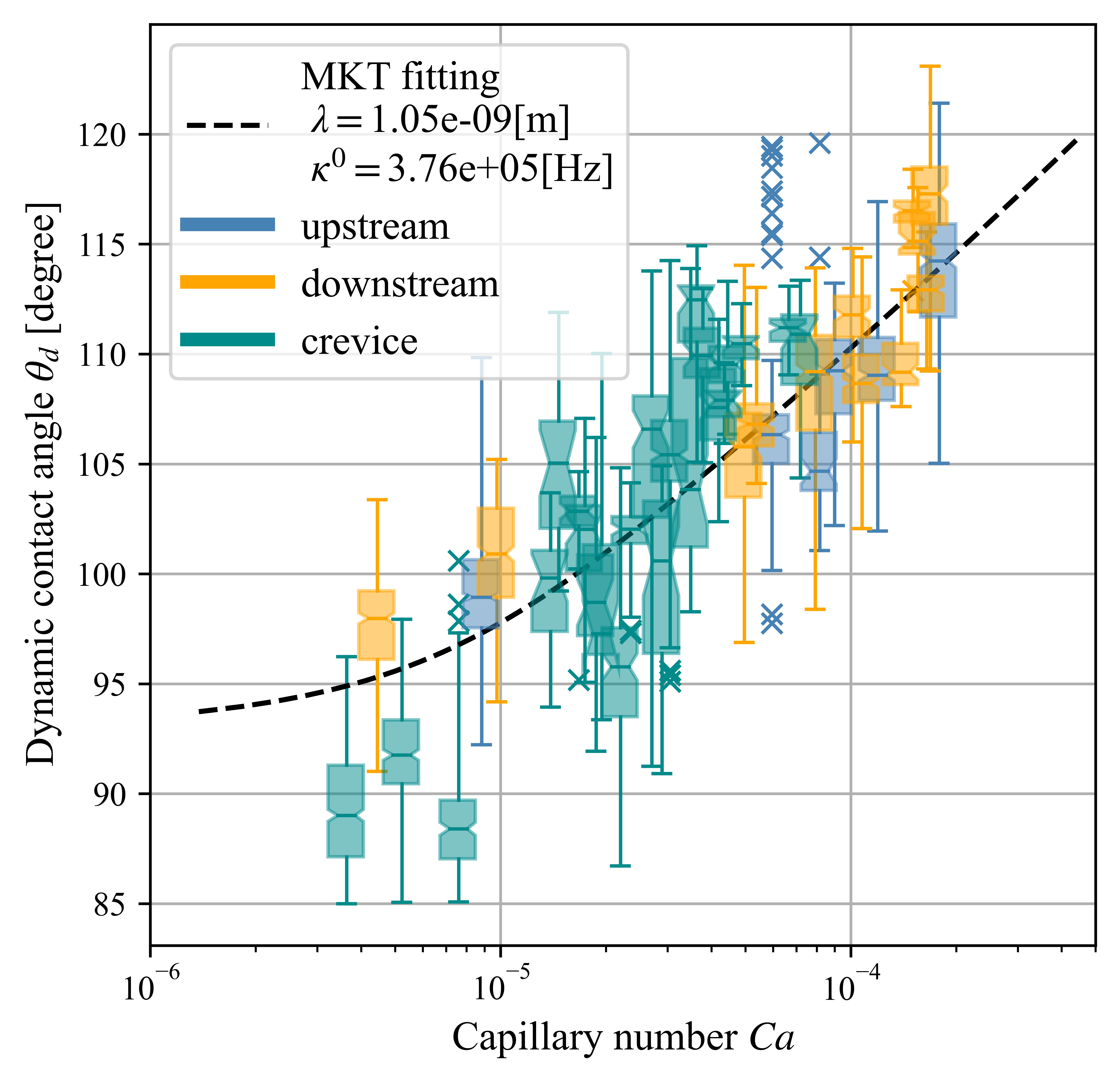}
		\caption{Areas of interest: ``upstream", ``downstream" and ``crevice"}
		\label{fig:3.3.1_dynCA}
	\end{subfigure}
	\hfill
	\begin{subfigure}[t]{.45\linewidth}
		\centering
		\includegraphics[width=\linewidth]{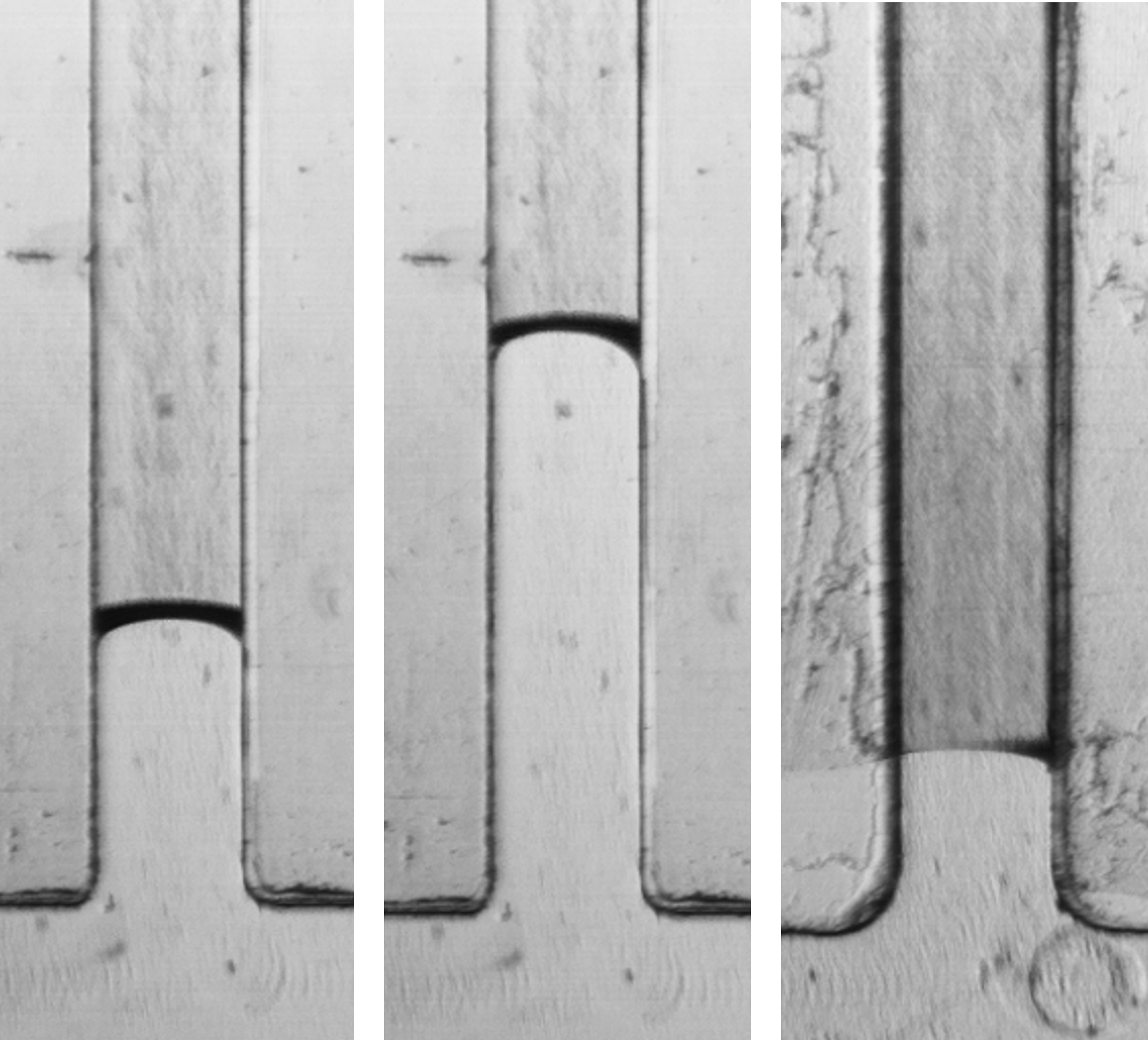}
		\caption{Examples of thick and asymmetrical interfaces in the ``crevice" area}
		\label{fig:3.3.2_dynCA_crevice}
	\end{subfigure}
	\caption{Dynamic contact angle $\theta_d$ versus capillary number $Ca$ with MKT fitting ($\lambda=\SI{1.13(3)}{nm}$, $\kappa^0=\SI{225.73(4358)}{kHz}$) as box plots. The lines inside the boxes mark the median and the interquartile range (IQR), indicated by the boxes, represents the middle 50\% (25\%-75\%) of the data. The whiskers extend to the lowest and highest data points within 1.5 times the IQR, with a few outliers plotted as ``x" individually beyond this range.}
	\label{fig:3.3_dynCA}
\end{figure}

To further validate the efficiency of the image processing algorithm, the resulting CA measurements are compared to a scientifically proven CA model - MKT model
\begin{equation}
\theta_d = \cos^{-1} \left( \cos{\theta_0} - \frac{2 k_B T}{\sigma \lambda^2} \sinh^{-1} \left( {\frac{U}{2 \kappa^0 \lambda}} \right) \right),
\label{eq:3.1_MKT}
\end{equation}
where $\theta_d$ is the dynamic CA, $k_B$ is the Boltzmann constant, $T$ is the temperature and $U$ is the interface velocity. The molecular motion frequency at equilibrium $\kappa^0$ and the average distance between active sites $\lambda$ are treated as free parameters and can be estimated via applying curve-fitting on the experimental data. Generally, $\lambda$ is in the order of molecular dimensions for small molecules from  \AA \ to \SI{}{nm} and $\kappa^0$ varies within several orders of magnitude from \SI{}{kHz} to \SI{}{GHz} as reported by \cite{Duvivier2013} and \cite{SEDEV2015661}.

The CL velocities are assumed to equal the interface motion velocities determined in \cref{subsec:3.1_flowRate}. Figure \ref{fig:3.3.1_dynCA} compares the dynamic CA from experiments and MKT fitting results as box plots, where a great agreement is visible with physically reasonable parameters. The curve fitting applied on the median values is realized using Python with the SciPy library developed by \cite{2020SciPy-NMeth}. The dynamic CA shows a strong dependency on the CL capillary number $Ca$, and the median values increase by around 20 degrees in the investigated $Ca$ range. A larger CA deviation in the ``crevice" area can be observed compared to the ``upstream" and ``downstream" regions, which might be attributed to the smaller width. The small dimension of the crevice could amplify the visualization of the local effects, leading to a more frequent interface pinning and thicker interface, as shown in \cref{fig:3.3.2_dynCA_crevice}. Furthermore, the occurrence of leakage (water infiltration between the plates) changes the wetting properties of channel walls and exhibits with the passing interface more of a $90 \degree$ angle (the rightmost crevice of \cref{fig:3.3.2_dynCA_crevice}), especially at small flow rates.

\subsection{Penetration depth}\label{subsec:3.3_penetrationDepth}

\begin{figure}[!htb]
	\centering
	\includegraphics[width=\linewidth]{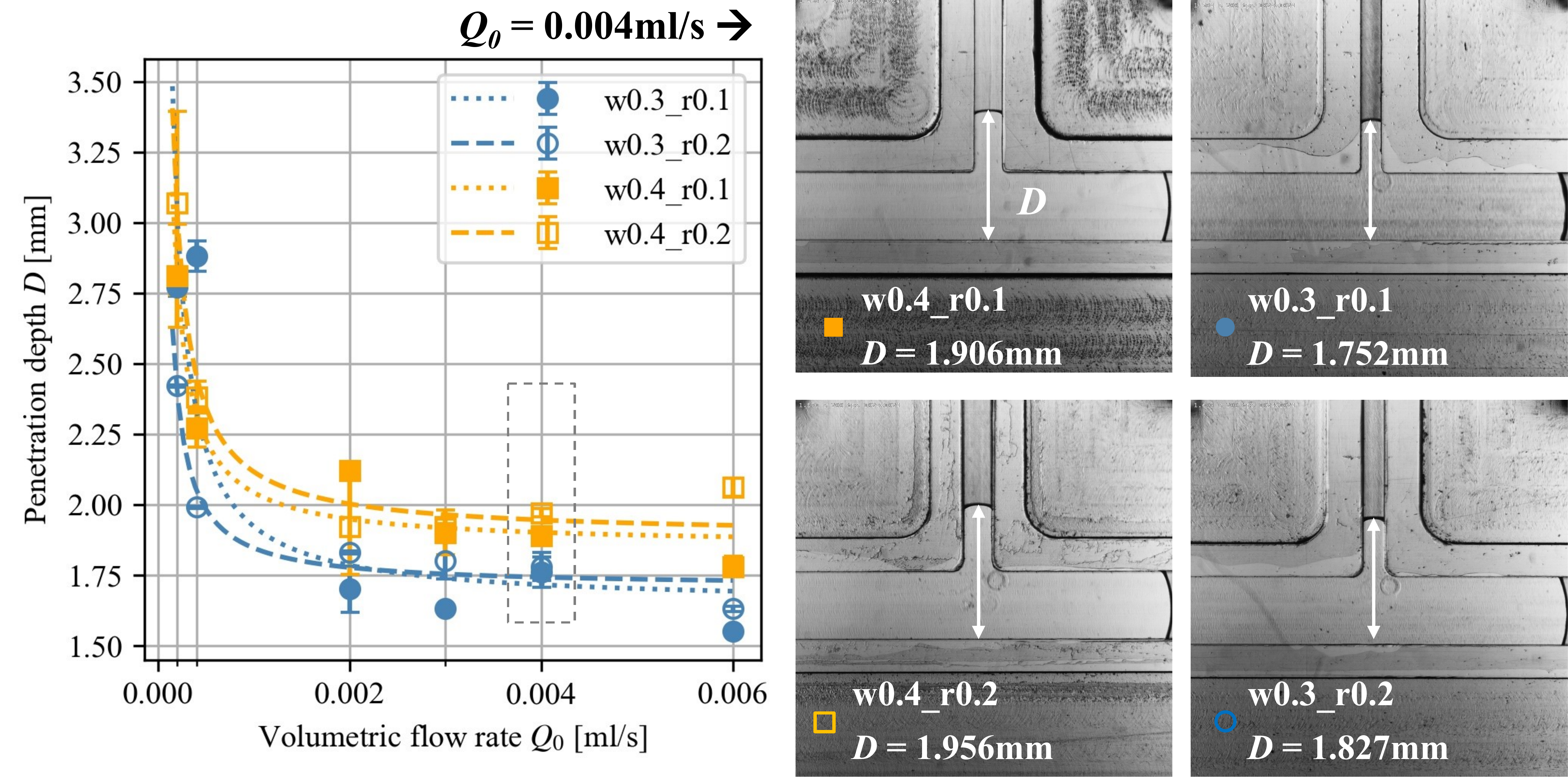}
	\caption{Left: Penetration depth $D$ in the crevice versus volumetric flow rate $Q_0$. Right: $Q_0=\SI{0.004}{ml/s}$}
	\label{fig:3.3_depth2}
\end{figure}

The penetration depth of the interface in the crevice $D$ is defined as the distance between the interface in the crevice and the bottom wall of the main channel, when the interface in the main channel is just about to leave the ROI, as depicted in \cref{fig:3.3_depth2}. The left of \cref{fig:3.3_depth2} shows the decrease of the penetration depth $D$ with increasing flow rate $Q_0$ and decreasing crevice width $w$. A linear regression analysis between $1/Q_0$ and $D$ is performed using the least squares method to better understand the influence of channel geometry on the penetration depth. Regarding the effect of the rounding edge radius $r$, a larger $r$ tends to give rise to $D$, with the exception that for smaller $Q_0$, the correlation can be reversed. For example, the samples with crevice width $w=\SI{0.3}{mm}$ at $Q_0=\SI{0.0002}{ml/s}$ and $Q_0=\SI{0.0004}{ml/s}$. This phenomenon can be associated with the fluid infiltration at the joint of both plates due to the manufacturing tolerances, driven by capillary forces. Specifically, a larger $r$ and smaller $Q_0$ increase the pinning time of the interface top on the left crevice corner (discussed in \cref{subsubsec:3.4.2_dt}), thereby lead to the fluid infiltration into the joint more pronounced and reduce the penetration depth $D$. In general, a slowly advancing interface is more sensitive to local surface effects, leading to a more irregular interface movement and penetration behavior.

\subsection{Pinning effect}\label{subsec:3.4_pinning}


Due to the T-junction, the flow is divided into two directions with decreased velocities: one continues the main channel, while the other flows into the orthogonal crevice. It is worth mentioning that due to the rounding radius, ``pinning" in this work does not imply that the CL remains completely stationary. Instead, here it means that the interface moves much slower along the rounded edge compared to its movement before and after passing through the T-junction.

\begin{figure}[!htb]
	\centering
	\includegraphics[width=0.9\linewidth]{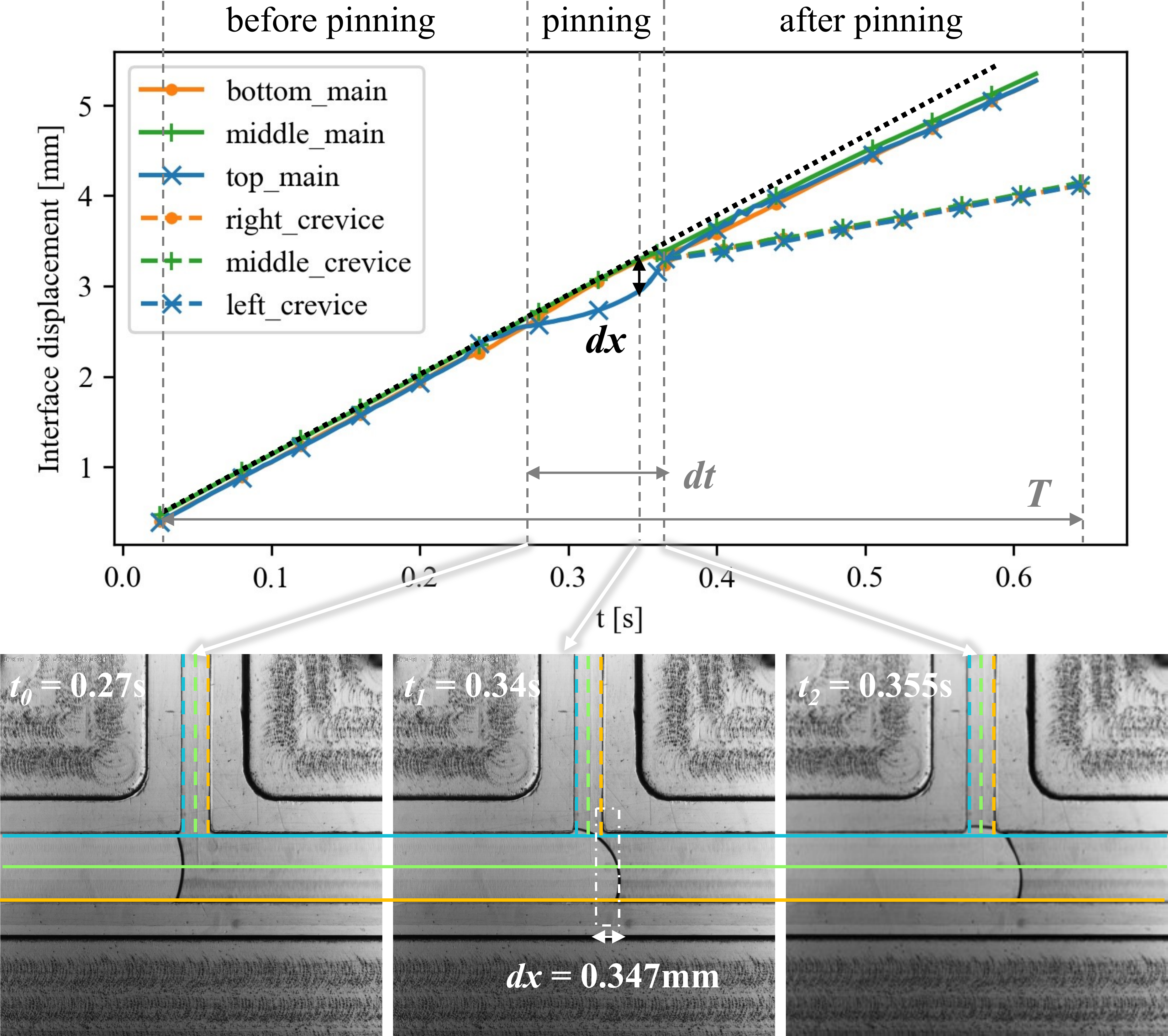}
	\caption{Top: Interface movement over time with $w=\SI{0.4}{mm}$ and $r=\SI{0.1}{mm}$ at $Q_0=\SI{0.004}{ml/s}$. Bottom: Start of pinning at $t=t_0$, maximum pinning distance $dx$ at $t=t_1$, end of pinning at $t=t_2$.}
	\label{fig:3.4_interface_movement}
\end{figure}

Figure \ref{fig:3.4_interface_movement} illustrates the interface movement over time along the x-axis and y-axis with $w=\SI{0.4}{mm}$, $r=\SI{0.1}{mm}$ and $Q_0=\SI{0.004}{ml/s}$, which can be classified as three stages: before pinning, pinning and after pinning. At the ``before pinning" stage, the interface advances steadily towards the T-junction. The positions of the interface top and bottom overlap in \cref{fig:3.4_interface_movement} due to the symmetrical interface shape, while the interface center is slightly ahead. At $t_0=\SI{0.27}{s}$, the interface top arrives at the left wall of the crevice and starts ``pinning" at the rounded corner, whereas the middle and bottom keep moving with a decreased velocity. This pinning effect leads to a maximum distance $dx=\SI{0.347}{mm}$ between the interface top and bottom at $t_1=\SI{0.34}{s}$, which occurs just before the interface top reaches the right wall of the crevice. Afterwards, the interface contacts the right corner of the crevice at $t_2=\SI{0.355}{s}$, causing the flow $Q_0$ to split into two directions as $Q_1$ and $Q_2$. Hence, $dt=t_2 - t_0=\SI{0.085}{s}$ refers to the ``pinning" time of the interface top on the left rounded corner. The interface movement over time in the y-direction is displayed on the right side of \cref{fig:3.4_interface_movement}. It begins from the end time of pinning at $t_2=\SI{0.355}{s}$ with a much smaller velocity compared with ``before pinning" stage. Moreover, due to the flow splitting and reduced flow rates, the slope of the interface movement in the ``after pinning" phase should go through a minor decline in comparison with ``before pinning" phase as well and can be observed in \cref{fig:3.4_interface_movement}. The black dotted line shows the movement of the interface middle position in ``before pinning" phase. The plots for two different flow rates are presented in \cref{fig:3.7_movements}. The time $T$ denotes the required time for the interface to flow through the main channel in ROI.

As quantitative means of the pinning effect, three parameters - maximum pinning distance $dx$, pinning time $dt$ and normalized pinning time $dt/T$ are employed and further investigated in the following subsections.

%
\subsubsection{Maximum pinning distance $dx$}\label{subsubsec:3.4.1_dx}

\begin{figure}[!htb]
	\centering
	\includegraphics[width=\linewidth]{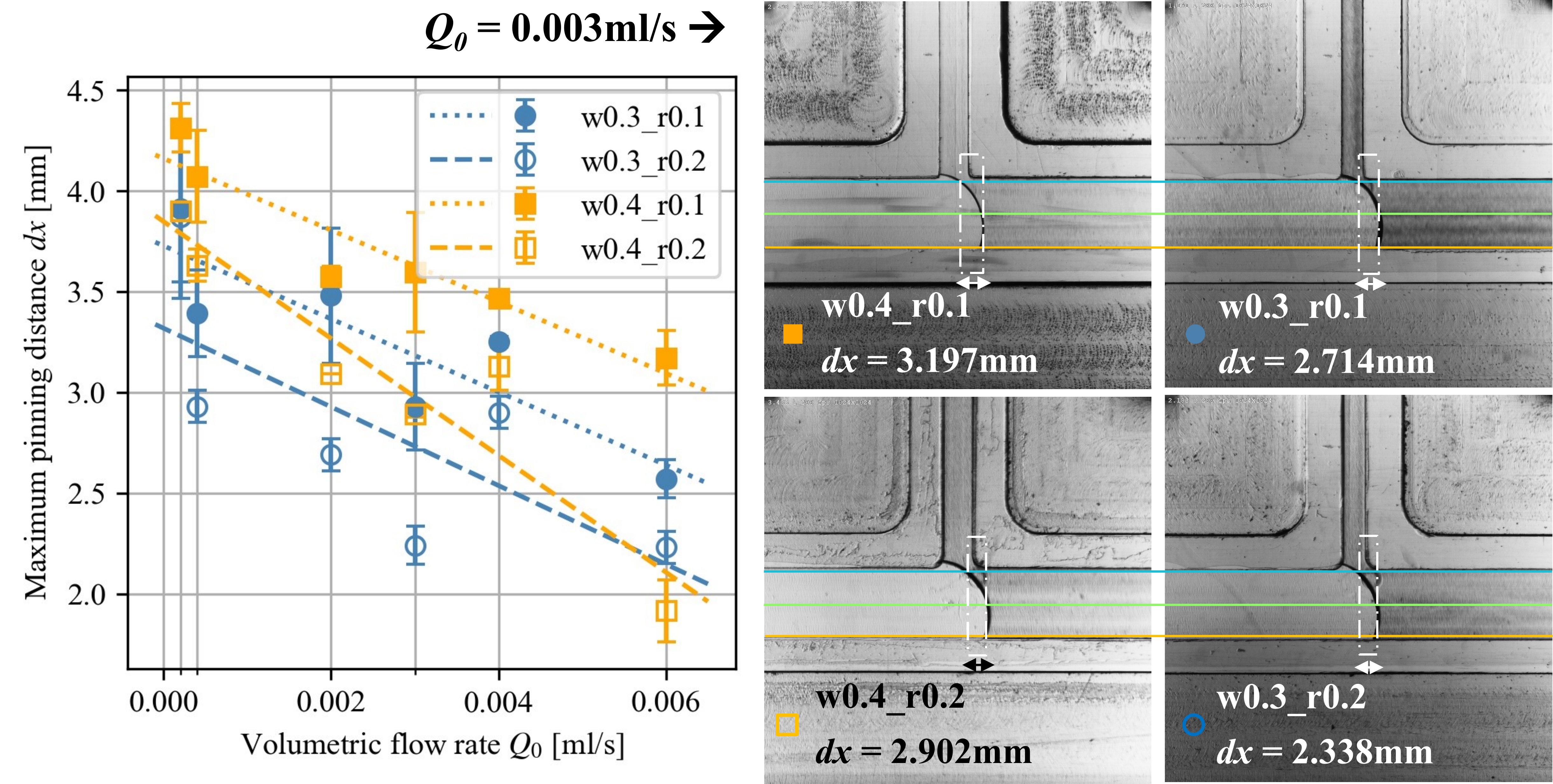}
	\caption{Left: Maximum pinning distance $dx$ versus volumetric flow rate $Q_0$. Right: $Q_0=\SI{0.003}{ml/s}$.}
	\label{fig:3.5_dx}
\end{figure}

In the left plot of \cref{fig:3.5_dx}, the maximum pinning distance $dx = max_{t \in T}(x_{bottom}(t)-x_{top}(t))$ exhibits an overall decreasing trend with increasing $Q_0$. This behavior is owing to the larger advancing dynamic CA and the resulting interface curvature, leading to a reduced maximum distance $dx$ between the interface top and bottom. Furthermore, a larger rounding radius $r$ allows the interface to progress further along the left wall of the crevice during pinning, resulting in the interface top advancing further ahead and thus a smaller $dx$. On the contrary, with a larger crevice width $w$, the $dx$ increases. Interestingly, the influence of a larger rounding $r$ appears to be more significant in decreasing the $dx$ compared with the effect of a smaller $w$. These trends are more apparent with the overlaid linear fits between $Q_0$ and $dx$ in \cref{fig:3.5_dx}. The right-hand side of \cref{fig:3.5_dx} presents the images of the interface with $dx$ marked for an identical flow rate $Q_0=\SI{0.003}{ml/s}$, providing a visual confirmation of the relationship between the geometric parameters $w$, $r$ and $dx$.

\subsubsection{Pinning time $dt$ and normalized pinning time $dt/T$}\label{subsubsec:3.4.2_dt}

\begin{figure}[!htb]
	\centering
	\begin{subfigure}[t]{0.48\linewidth}
		\centering
		\includegraphics[width=\linewidth]{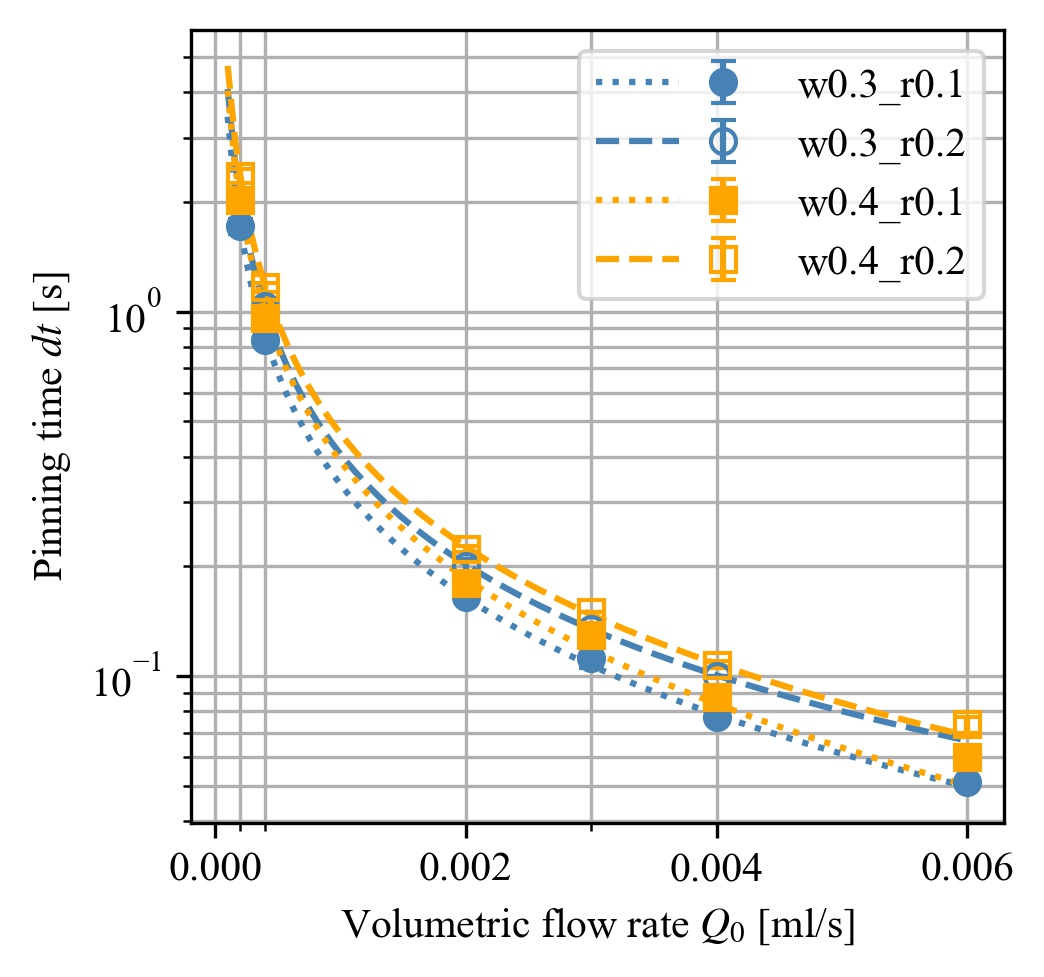}
		\caption{Pinning time $dt$.}
		\label{fig:3.6.1_dt}
	\end{subfigure}
	\hfill
	\begin{subfigure}[t]{0.48\linewidth}
		\centering
		\includegraphics[width=\linewidth]{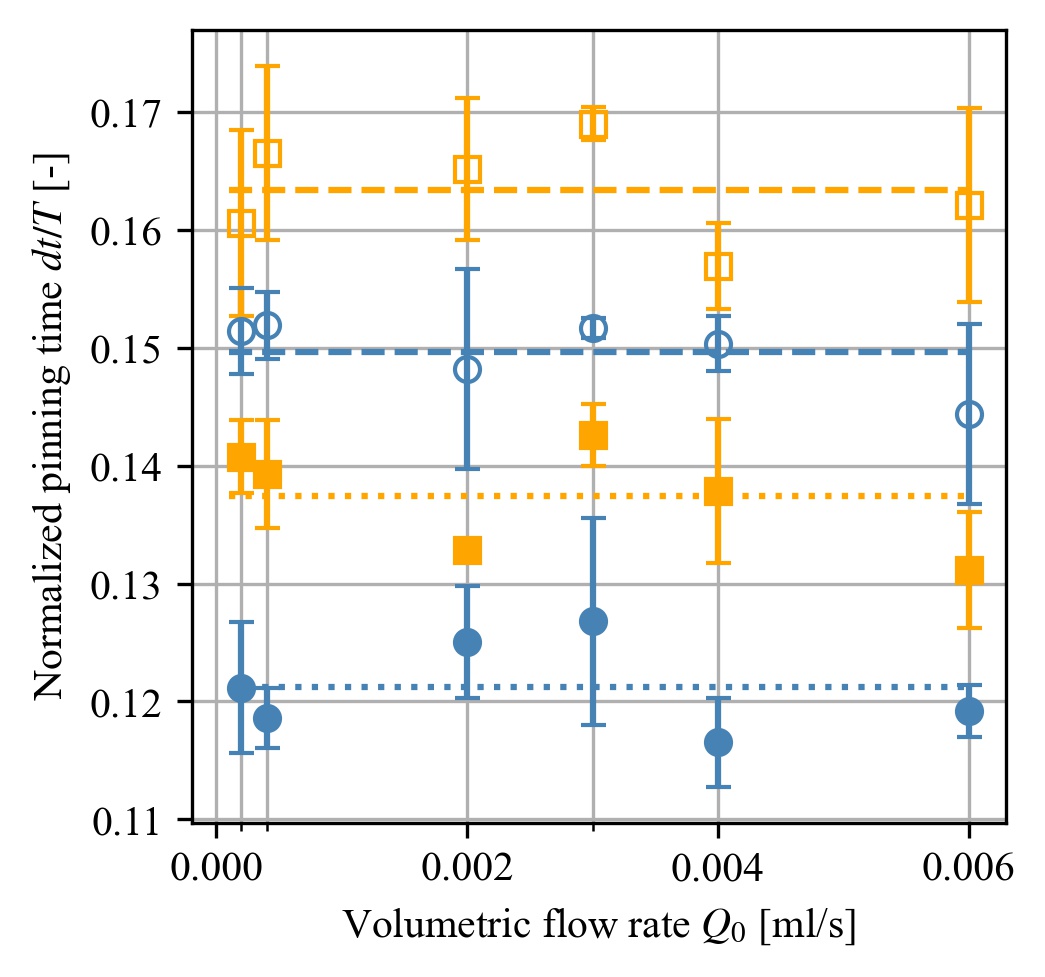}
		\caption{Normalized pinning time $dt/T$.}
		\label{fig:3.6.2_dt_t}
	\end{subfigure}
	\hfill
	\begin{subfigure}[t]{0.8\linewidth}
		\includegraphics[width=\linewidth]{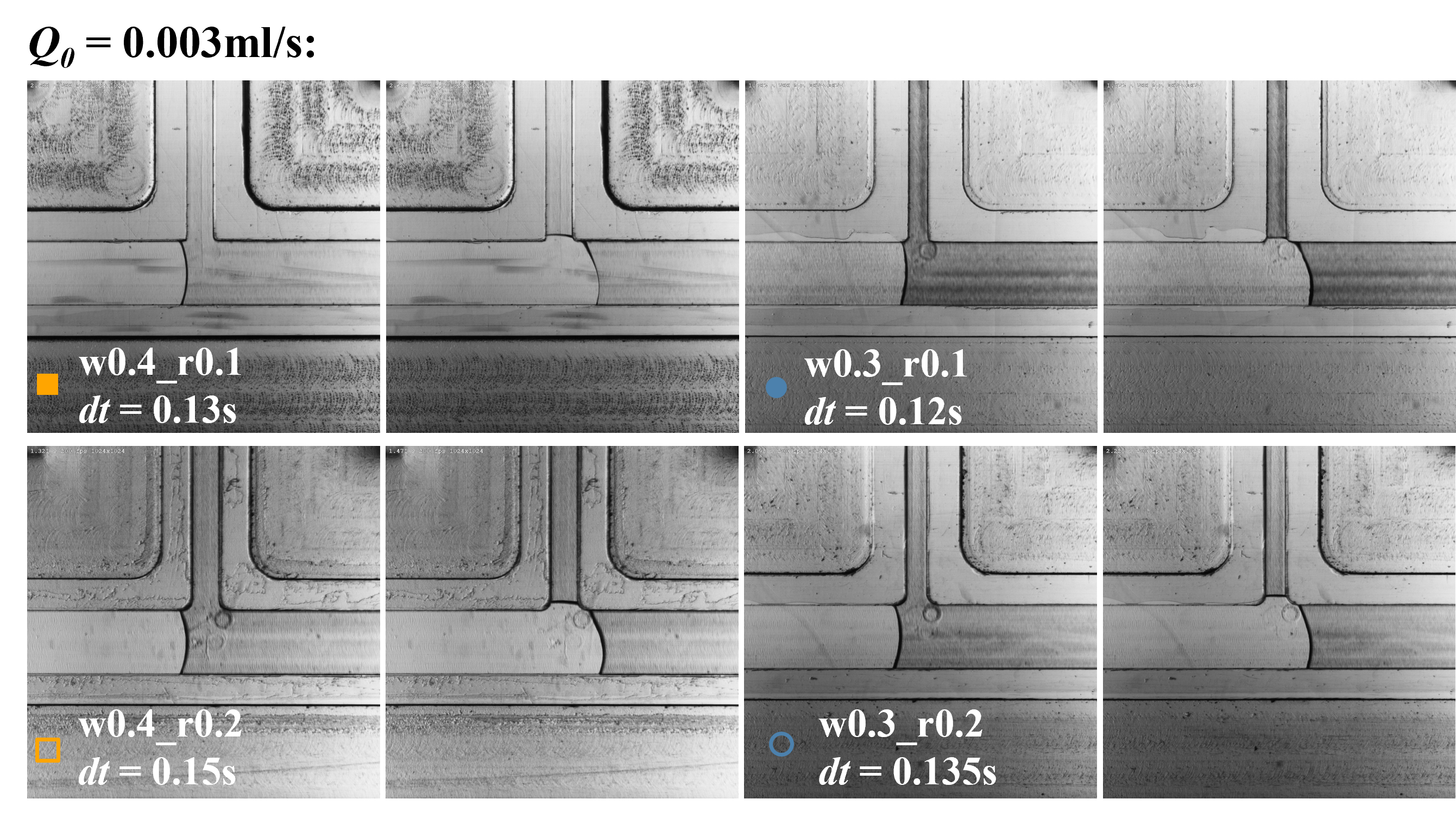}
		\caption{Pinning time $dt$ at $Q_0=\SI{0.003}{ml/s}$ for different crevice widths and rounding edge radii.}
		\label{fig:3.6.3_dtQ0}
	\end{subfigure}
	\caption{Interface pinning time $dt$ and normalized pinning time $dt/T$ for different volumetric flow rates $Q_0$.}
	\label{fig:3.6_dt}
\end{figure}

In addition to the maximum pinning distance $dx$, the interface pinning time $dt=t_2 - t_0$ is analyzed, as displayed in \cref{fig:3.6.1_dt}. The start time of pinning $t_0$ is defined as the time at which the distance between the interface top and bottom reaches or exceeds 30 pixels $|x_{top}-x_{bottom}| \geqq 30 \ \text{pixels}$, and this distance continues to increase for the last three frames. This criterion helps to exclude the effect of a potential asymmetric interface shape. The end time of pinning $t_2$ is defined as the time at which dark pixels appear in the ``y-axis" area along the right interface tracking line (orange dashed line in \cref{fig:3.4_interface_movement}).

Similar to $dx$, a narrower crevice and larger flow rate decrease the pinning time $dt$. The data analysis reveals the pinning time $dt$ to be inversely proportional to $Q_0$. This relationship is intuitive and aligns with the expectations. Figure \ref{fig:3.6.3_dtQ0} presents the start and end images of pinning time $dt$ of four samples at $Q_0=\SI{0.003}{ml/s}$. Interestingly, in contrast to the maximum pinning distance $dx$, a larger rounding radius tends to increase $dt$. This trend can be explained by the definition of $dt$, as mentioned before. It describes the duration during which the interface top starts ``pinning" at the left corner, leading to a deviation of the interface from its symmetrical motion, and reaches the right corner of the crevice at the end of ``pinning". With a larger rounding radius $r$, while having the same crevice width $w$, the interface pins earlier at the T-junction due to the geometrical conditions and contacts the right corner later, resulting in a longer pinning time. However, this observation does not indicate an enhanced ``pinning" effect by the rounding geometry but rather diminished, as evidenced by the evaluation of penetration depth in \cref{fig:3.3_depth2} and maximum pinning distance in \cref{fig:3.5_dx}. 

In addition to the pinning time $dt$, the normalized pinning time $dt/T$ is plotted in \cref{fig:3.6.2_dt_t}. Fitting the data reveals a linear correlation between $dt/T$ and the inflow rate $Q_0$. Similar to $dt$, a larger crevice width $w$ and rounding radius $r$ lead to an increase in $dt/T$, while it remains independent of $Q_0$. Another noteworthy observation is that both $dt$ and $dt/T$ appear to be more sensitive to the rounding radius $r$ than to the crevice width $w$. Furthermore, to better represent the effect of the channel geometry on the fluid penetration behavior, a non-dimensionless parameter $w/r$ as the ratio between the crevice width and the rounding edge radius is evaluated. However, as a result of the lack of a clear and significant impact of the ratio $w/r$ on most of the investigated parameters, the plots are included in \cref{appendice02}.

\section{Summary and conclusion}\label{sec:04_conclusion}

A comprehensive understanding of the wetting and penetration characteristics is essential in reliable product design, particularly regarding sealing performance and corrosion resistance. This work experimentally investigates the geometrical effects on fluid wetting and penetration behaviors in a T-shaped microchannel, consisting of a main channel and an orthogonally placed narrower crevice. The robust and reproducible experimental images are evaluated with an automated image processing method, providing an accurate and straightforward approach to determine the interface motion and the CA at the wall. The algorithm allows to analyze large datasets and reduce the need for manual measurements. Through a systematic evaluation of the results, two main aspects are presented and discussed: the fluid penetration into the crevice and the interface dynamic within the T-junction. The interpretation is based on the results of the automated image processing algorithm, which provides the interface position within the channel and the CA at each time. The algorithm undergoes a sanity check for flow rate conservation and the CAs determined are compared to a scientifically proven CA model. Both show excellent agreement, paving the path for the following evaluations.

The penetration depth $D$ of the interface in the crevice, determined by addressing the interface movement in the ``y-axis" area, decreases with smaller crevice width and rounding radius. The data analysis shows an inversely proportional behavior of the penetration depth to the volumetric inflow rate. In terms of the temporal evolution of interface displacement in the ``x-axis" area, it can be classified into three stages: before pinning, pinning and after pinning, where the maximum pinning distance $dx$, pinning time $dt$ and normalized pinning time $dt/T$ are quantified and compared. The first two parameters are shown to decrease with larger flow rates and smaller crevice width, while $dt/T$ remains constant with varying flow rates. On the contrary, $dx$ is inversely proportional to the rounding radius, while $dt$ shows a positive correlation. The rounding radius $r$ contributes more to the pinning effect than the crevice width $w$. This fact implies that introducing geometric features such as grooves or bends can be more efficient than simply narrowing the sealing joint when aiming for a more reliable product design in terms of sealing performance.

The presented results shed new light on the dynamic wetting and penetration behavior in microchannels with complex geometries from an industrial application perspective. The measured data also allows the determination of the CL dynamics within the parameter space. This enables later numerical investigation by providing the fitted MKT model as a wall boundary condition. In the next step, a numerical framework is to be established along with additional experimental work, to create a large regime map, categorizing a more diverse parameter set in rounding radii, channel widths and materials. Based on data science, we hope to provide correlations that guide future engineers in designing better joint geometries and positions and improving sealing performance. Moreover, as suggested, additional experiments involving other fluids with varying viscosities and surface energies will be conducted. This broader investigation will enhance the applicability of the findings, providing more extensive insights into fluid wetting characteristics across different scenarios and contributing to the development of more robust designs.

\appendix
\section{Appendix A: Interface displacement over time}\label{appendice01}

\begin{figure}
	\centering
	\includegraphics[width=\linewidth]{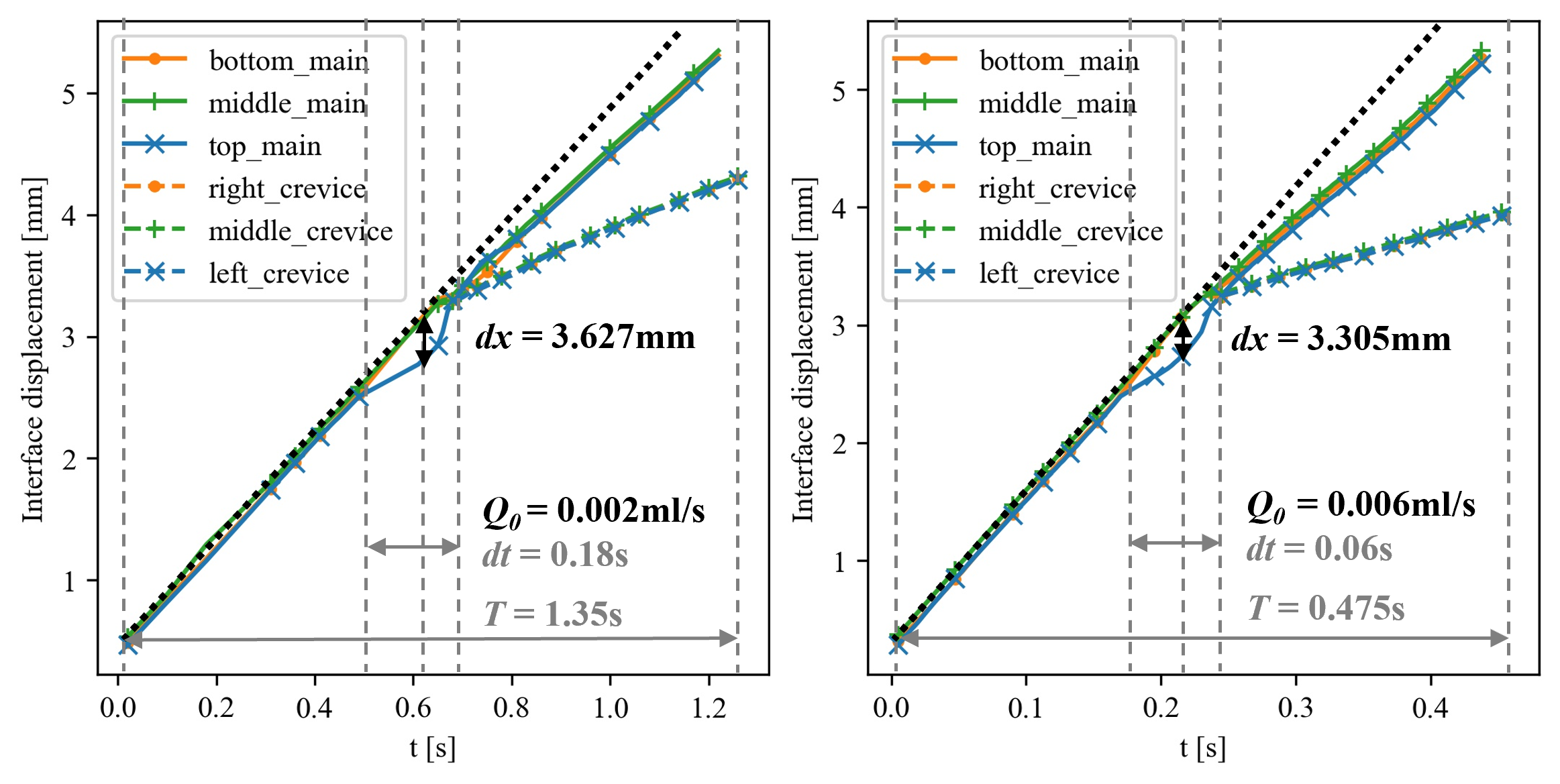}
	\caption{Interface movement over time of sample with $w=\SI{0.4}{mm}$ and $r=\SI{0.1}{mm}$ at  $Q_0=\SI{0.002}{ml/s}$ and $Q_0=\SI{0.006}{ml/s}$.}
	\label{fig:3.7_movements}
\end{figure}

The temporal evolution of interface displacement of sample with $w=\SI{0.4}{mm}$ and $r=\SI{0.1}{mm}$ at $Q_0=\SI{0.002}{ml/s}$ and $Q_0=\SI{0.006}{ml/s}$ is shown in \cref{fig:3.7_movements}. Similar to \cref{fig:3.4_interface_movement}, the interface movement in the main channel is plotted with that in the crevice superimposed on the right side, beginning from the end of the pinning stage. The results confirm that both the pinning time $dt$ and maximum pinning distance $dx$ decrease with increasing $Q_0$. Moreover, the decline in CL velocities after the T-junction, as a result of the flow splitting into two directions, is evident. This behavior is highlighted with the black dotted line as a reference for the change in the interface velocities.

\section{Appendix B: Non-dimensional parameter $w/r$}\label{appendice02}

\begin{figure}[!htb]
	\centering
	\begin{subfigure}[b]{.58\linewidth}
		\centering
		\includegraphics[width=\linewidth]{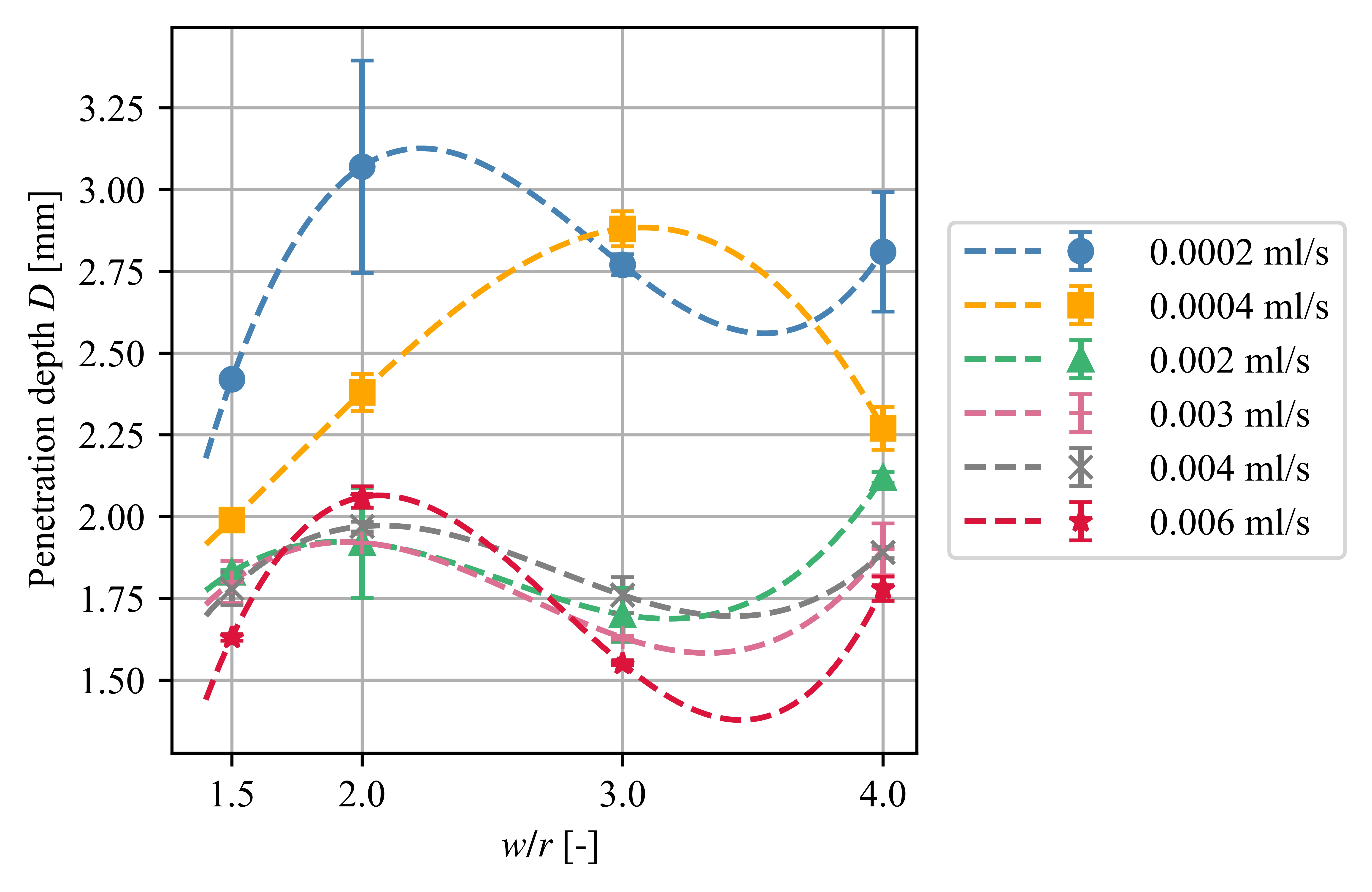}
	\end{subfigure}
	\hfill
	\begin{subfigure}[b]{.4\linewidth}
		\centering
		\includegraphics[width=\linewidth]{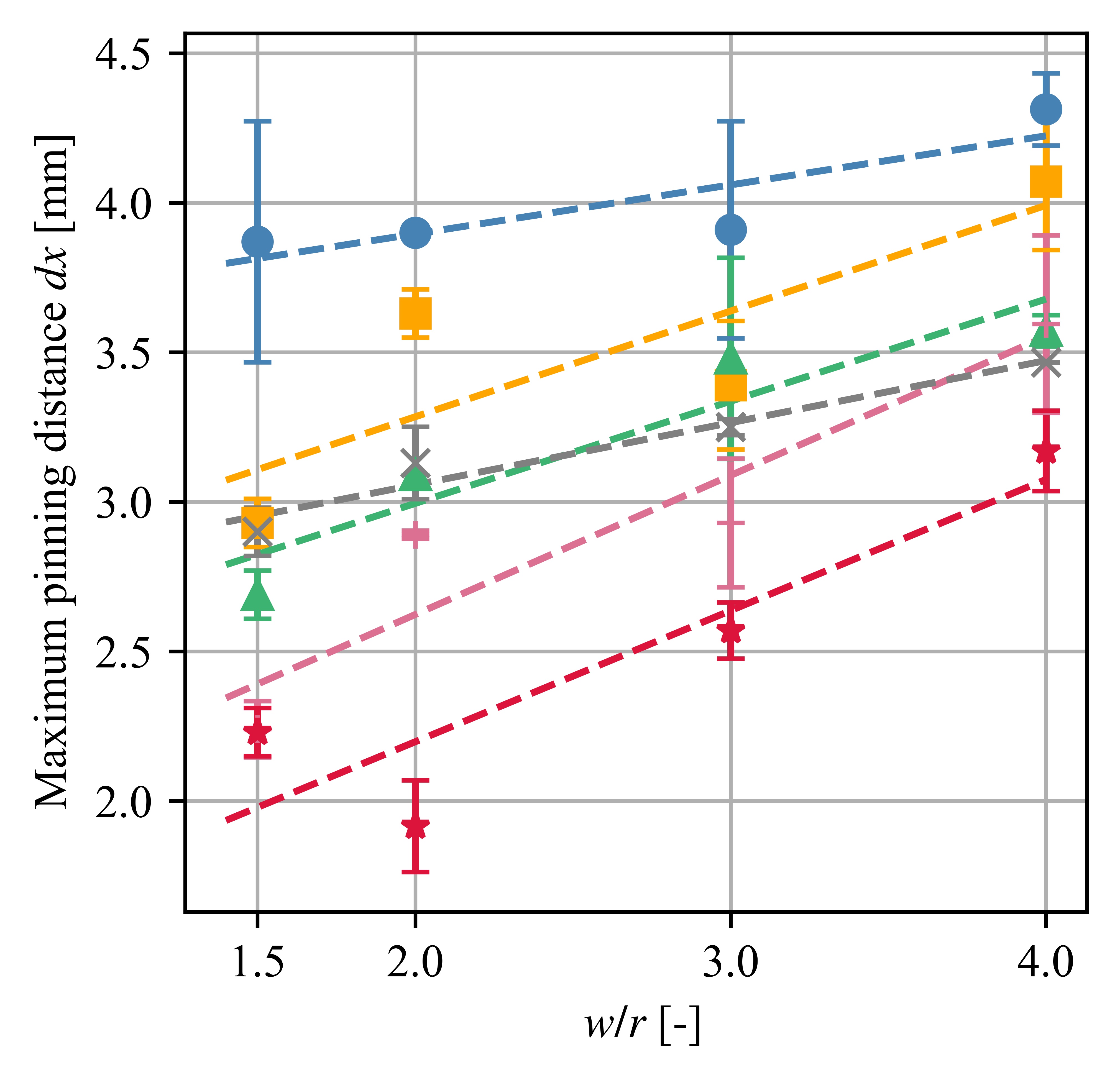}
	\end{subfigure}
	\hfill
	\begin{subfigure}[b]{.58\linewidth}
		\centering
		\includegraphics[width=\linewidth]{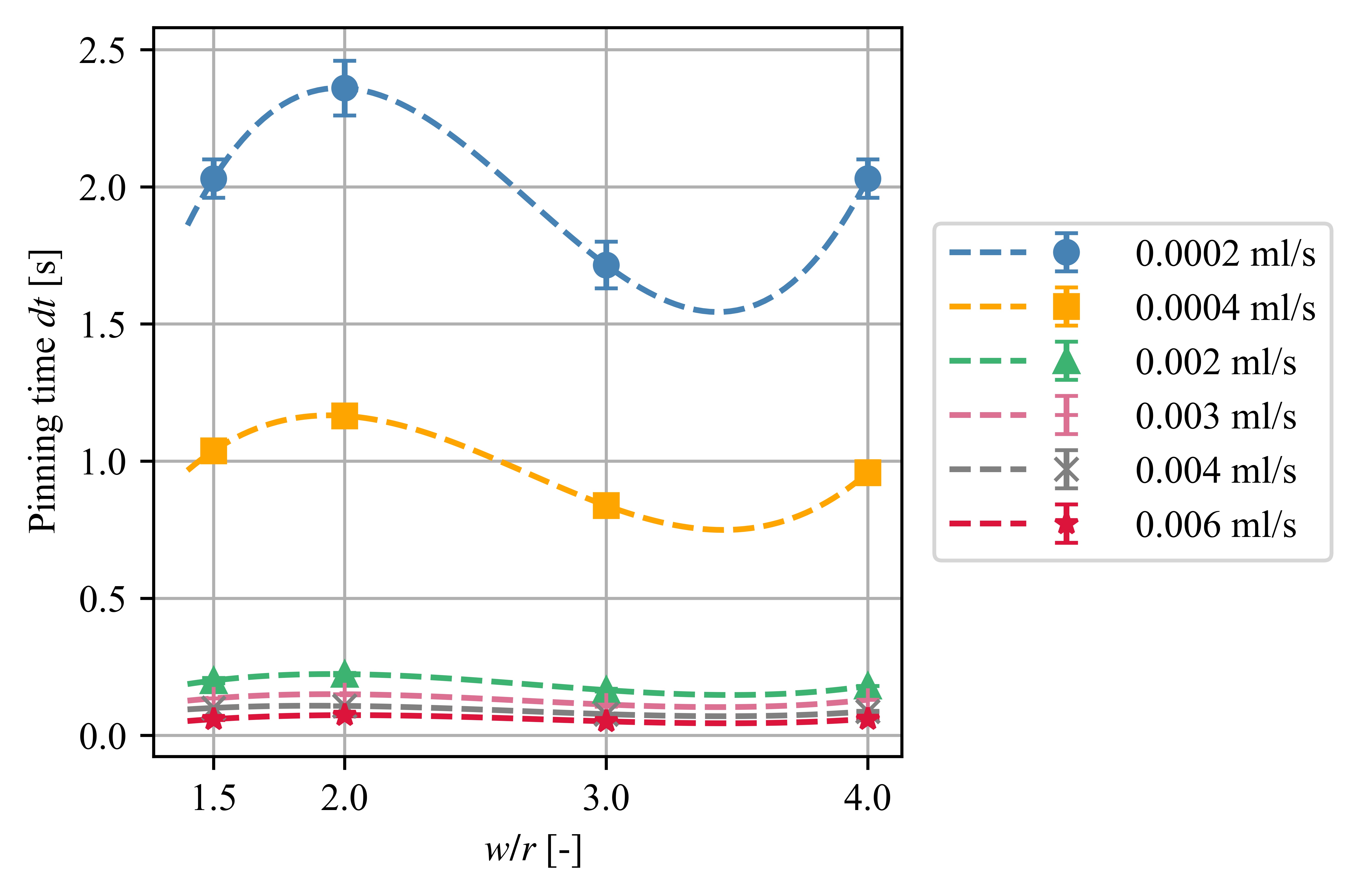}
	\end{subfigure}
	\hfill
	\begin{subfigure}[b]{.4\linewidth}
		\centering
		\includegraphics[width=\linewidth]{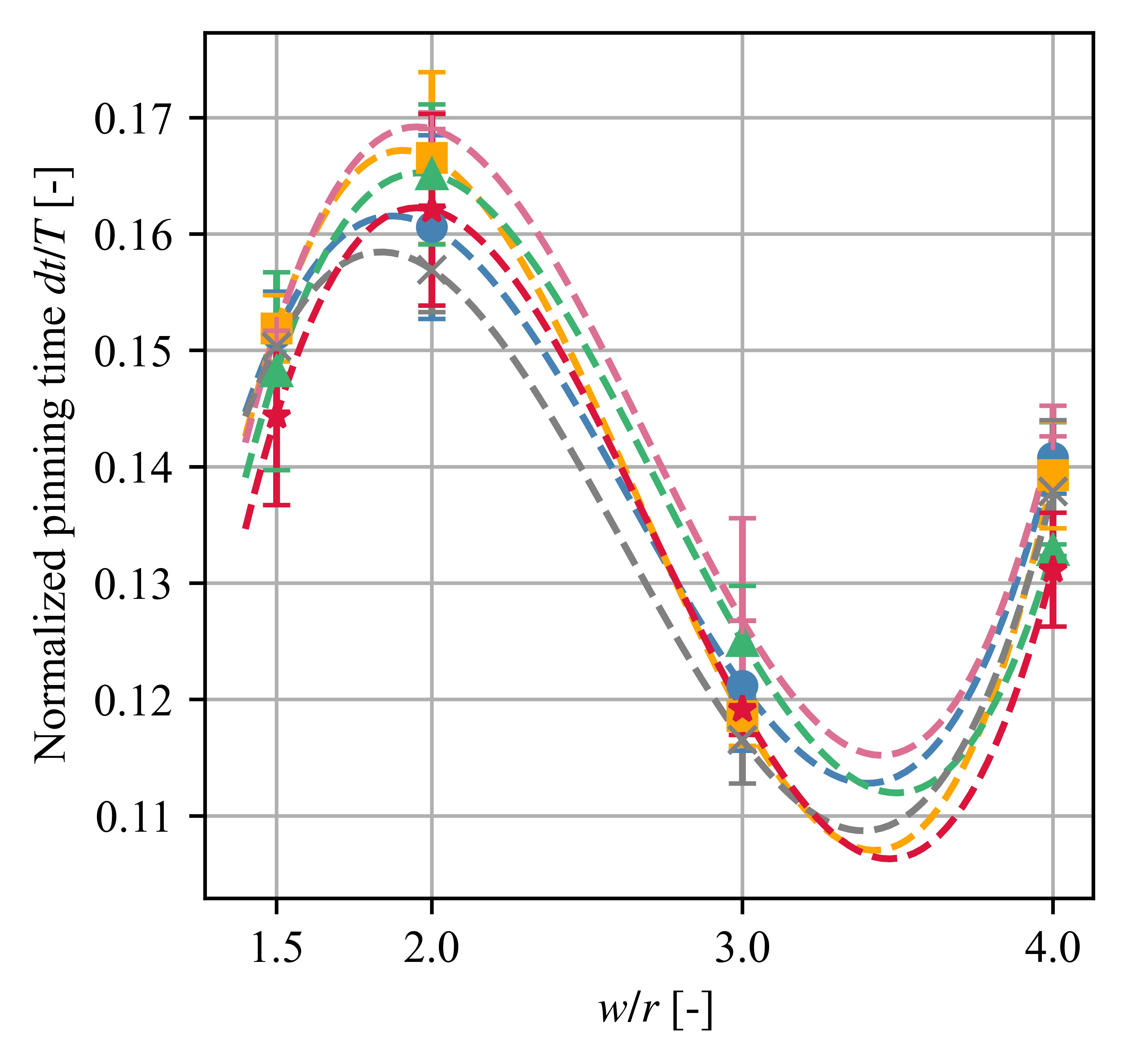}
	\end{subfigure}
	\caption{The penetration depth $D$, maximum pinning distance $dx$, pinning time $dt$ and the normalized pinning time $dt/T$ versus the ratio between the crevice width and the rounding radius $w/r$.}
	\label{fig:wr}
\end{figure}

A non-dimensionless parameter $w/r$ as the ratio between the crevice width and the rounding edge radius is evaluated, as listed in \cref{tab:wr}. The results are presented in \cref{fig:wr}, with polynomial fits applied to better visualize the impact of the ratio $w/r$. It should be noted that these polynomial fits are only intended for visualization purposes and do not hold any physical significance. While a positive linear relationship between the maximum pinning distance $dx$ and $w/r$ is apparent, the other three parameters $D$, $dt$ and $dt/T$ do not exhibit a strong correlation with this ratio $w/r$. Nevertheless, this aspect $w/r$ is to be incorporated into the future work by including a broader range of channel geometries.

\begin{table}[!htb]
	\centering
	\begin{tabular}{c c c c}
		\toprule[1.5pt]
		Sample & $w$ (\SI{}{mm}) & $r$ (\SI{}{mm}) & $w/r$ \\
		\midrule[0.5pt]
		w0.4\_r0.1 & 0.4 & 0.1 & 4 \\
		w0.4\_r0.2 & 0.4 & 0.2 & 2 \\
		w0.3\_r0.1 & 0.3 & 0.1 & 3 \\
		w0.3\_r0.2 & 0.3 & 0.2 & 1.5 \\
		\bottomrule[1.5pt]
	\end{tabular}
	\caption{The test sample variations with crevice width $w$, rounding edge radius $r$ and the ratio $w/r$.}
	\label{tab:wr}
\end{table}

\section*{Acknowledgments}
The authors would like to thank Dr. Mathis Fricke of Technical University of Darmstadt for the helpful and valuable discussion on the experimental results.
The last author acknowledges the funding by the German Research Foundation (DFG): July 1 2020 - 30 June 2024 funded by the German Research Foundation (DFG) - Project-ID 265191195 - SFB 1194.

\bibliography{sn-bibliography}

\end{document}